\documentclass[conference]{IEEEtran}
\usepackage{cite}
\usepackage{amsmath,amssymb,amsfonts}
\usepackage{algorithmic}
\usepackage{graphicx}
\usepackage{textcomp}
\usepackage{xcolor}
\usepackage[bookmarks=false,hidelinks]{hyperref}
\usepackage{booktabs}
\usepackage{multirow}
\usepackage{nicefrac}
\usepackage{enumitem}
\usepackage{listings}
\usepackage{pgfplots}
\usepgfplotslibrary{statistics}
\pgfplotsset{compat=1.14}
\usepackage[framemethod=tikz]{mdframed}
\def\BibTeX{{\rm B\kern-.05em{\sc i\kern-.025em b}\kern-.08em
    T\kern-.1667em\lower.7ex\hbox{E}\kern-.125emX}}

\mdfdefinestyle{mystyle}{%
  rightline=true,
  innerleftmargin=10,
  innerrightmargin=10,
  outerlinewidth=2pt,
  topline=false,
  rightline=true,
  bottomline=false,
  skipabove=\topsep,
  skipbelow=\topsep,
  frametitlefont=\normalfont\sc
}
\DeclareFixedFont{\ttb}{T1}{txtt}{bx}{n}{8}
\DeclareFixedFont{\ttm}{T1}{txtt}{m}{n}{8}
\usepackage{color}
\definecolor{deepblue}{rgb}{0,0,0.5}
\definecolor{deepred}{rgb}{0.6,0,0}
\definecolor{deepgreen}{rgb}{0,0.5,0}
\newcommand\compilerstyle{\lstset{
  basicstyle=\ttm,
  keywordstyle=\ttb\color{deepblue},
  frame=tb,
  showstringspaces=false            % 
}}
\newcommand\pythonstyle{\lstset{
  language=Python,
  basicstyle=\ttm,
  otherkeywords={self},
  keywordstyle=\ttb\color{deepblue},
  emph={MyClass,__init__},
  emphstyle=\ttb\color{deepred},
  stringstyle=\color{deepgreen},
  frame=tb,
  showstringspaces=false            % 
}}
\lstnewenvironment{python}[1][]
{
\pythonstyle
\lstset{#1}
}
{}
\lstnewenvironment{compiler}[1][]
{
\compilerstyle
\lstset{#1}
}
{}
\newcommand\pythoninline[1]{{\pythonstyle\lstinline!#1!}}
\newcommand\compilerinline[1]{{\compilerstyle\lstinline!#1!}}

\begin{document}
\begin{sloppy}

\author{\IEEEauthorblockN{Emillie Thiselton and Christoph Treude}
\IEEEauthorblockA{School of Computer Science\\University of Adelaide\\
emillie.thiselton@gmail.com, christoph.treude@adelaide.edu.au}
}

\title{Enhancing Python Compiler Error Messages\\via Stack Overflow}
\IEEEoverridecommandlockouts
\IEEEpubid{\makebox[\columnwidth]{\hfill} \hspace{\columnsep}\makebox[\columnwidth]{ }}
\maketitle
\IEEEpubidadjcol

\begin{abstract}
Background: Compilers tend to produce cryptic and uninformative error messages, leaving programmers confused and requiring them to spend precious time to resolve the underlying error. To find help, programmers often take to online question-and-answer forums such as Stack Overflow to start discussion threads about the errors they encountered.

Aims: We conjecture that information from Stack Overflow threads which discuss compiler errors can be automatically collected and repackaged to provide programmers with enhanced compiler error messages, thus saving programmers' time and energy.

Method: We present Pycee, a plugin integrated with the popular Sublime Text IDE to provide enhanced compiler error messages for the Python programming language. Pycee automatically queries Stack Overflow to provide customised and summarised information within the IDE. We evaluated two Pycee variants through a think-aloud user study during which 16 programmers completed Python programming tasks while using Pycee.

Results: The majority of participants agreed that Pycee was helpful while completing the study tasks. When compared to a baseline relying on the official Python documentation to enhance compiler error messages, participants generally preferred Pycee in terms of helpfulness, citing concrete suggestions for fixes and example code as major benefits.

Conclusions: Our results confirm that data from online sources such as Stack Overflow can be successfully used to automatically enhance compiler error messages. Our work opens up venues for future work to further enhance compiler error messages as well as to automatically reuse content from Stack Overflow for other aspects of programming.
\end{abstract}


\section{Introduction}

\begin{figure*}
\centering
\includegraphics[width=\linewidth]{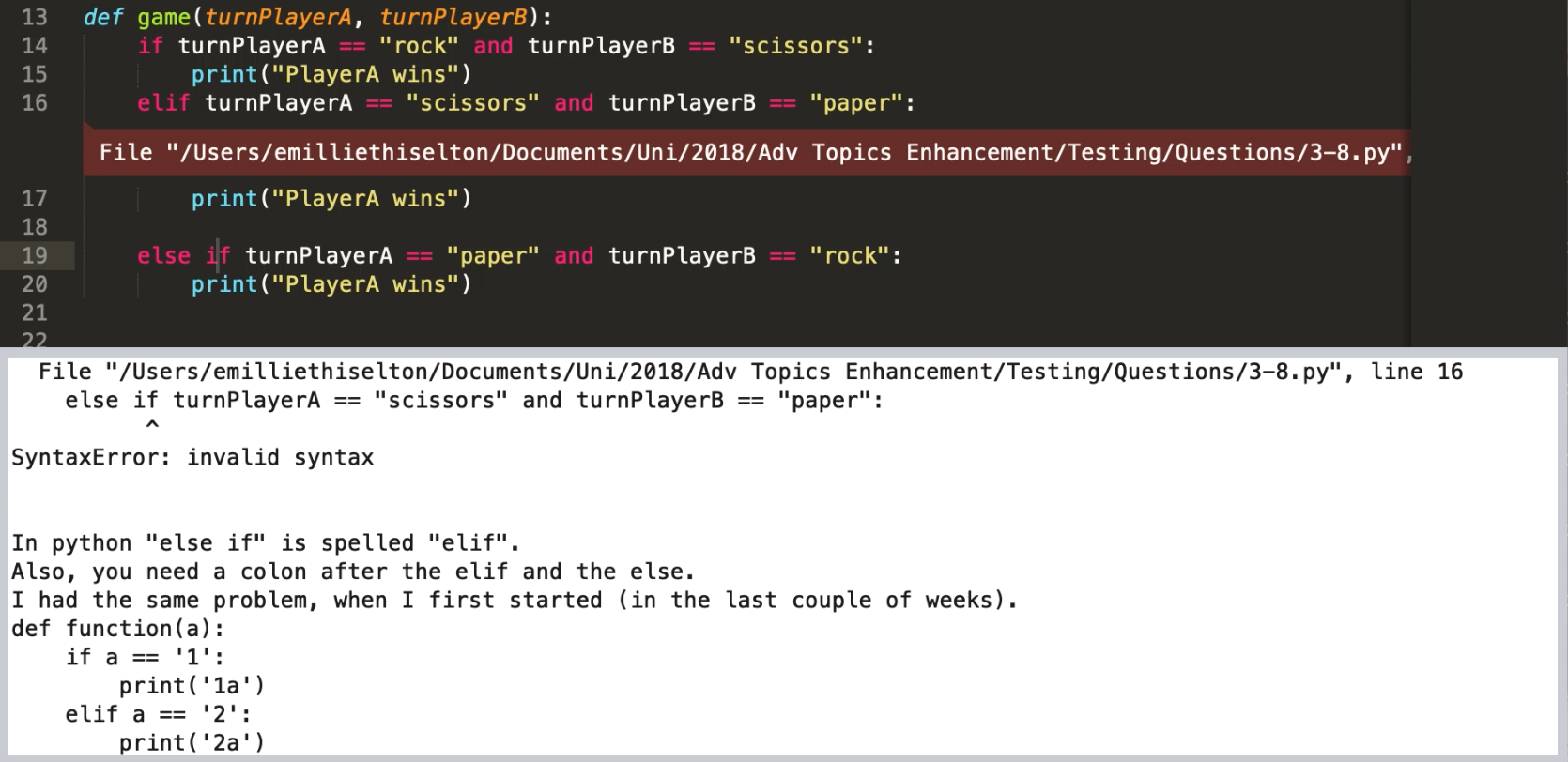}
\caption{Screenshot of \textsc{Pycee}. The first few lines on white background show the original compiler error message produced by Python, the additional lines show the enhanced error message produced by \textsc{Pycee}. The message provided by \textsc{Pycee} is a summary of Stack Overflow answer 2395167. Note that in the screenshot, the offending line has already been corrected.}
\vspace{-0.5cm}
\label{fig:pycee}
\end{figure*}

In Richard L.~Wexelblat's 1976 paper ``Maxims for malfeasant designers, or how to design languages to make programming as difficult as possible''~\cite{Wexelblat1976}, published at the second International Conference on Software Engineering, the author facetiously proposes ``cryptic diagnostics'' as one way of maximising difficulty for the user, arguing that a useless compiler error message should only state the effect of an error instead of its cause. Unfortunately, the reality is often not far from this tongue-in-cheek proposal, and not much has changed in terms of the helpfulness of compiler error messages in the meantime: Programmers often encounter cryptic compiler error messages that are difficult to understand and thus difficult to resolve~\cite{Traver2010}, compiler error messages are cryptic, uninformative, terse, and misleading and pose a significant barrier to progress~\cite{Becker2016a}, and they fail to convey information accurately~\cite{Marceau2011}.

When faced with cryptic error messages, Wexelblat suggested that ``at best, the programmer can get more information from a manual. More often, there is no help available''. This aspect has changed significantly since the mid-70s with the availability of software documentation from a multitude of sources on the Internet. For example, on the question-and-answer forum Stack Overflow,\footnote{\url{https://stackoverflow.com/}} almost 15,000 questions have been tagged with \compilerinline{compiler-error},\footnote{\url{https://stackoverflow.com/questions/tagged/compiler-errors}} with many more questions requesting help in understanding and addressing a compiler error without the explicit tag---a simple search for \compilerinline{compiler error} on Stack Overflow returns more than 350,000 results.\footnote{\url{https://stackoverflow.com/search?q=compiler+error}} Past work has found that questions which include a specific error message are the fourth most common question type on the site~\cite{Treude2011} and that debug-corrective questions are common~\cite{deSouza2014, Sillito2012}. 

Following work on automatically combining data from multiple sources related to software development with the ultimate goal of making programmers more productive (e.g.,~\cite{Delfim2016, Chatterjee2017, Ponzanelli2014a, Treude2016}), in this paper, we introduce \textsc{Pycee} (\textbf{Py}thon \textbf{C}ompiler \textbf{E}rror \textbf{E}nhancer), a plugin for the popular Sublime Text IDE\footnote{\url{https://www.sublimetext.com/}} which automatically augments Python compiler error messages with data from the question-and-answer website Stack Overflow to provide programmers with additional information in concise and summarised form and to offer concrete suggestions for error resolution. We chose to target Python since it is the fastest-growing major programming language today, edging out Java according to the recently published results of the 2019 Stack Overflow Developer Survey.\footnote{\url{https://insights.stackoverflow.com/survey/2019}} In addition, other programming languages, such as Java~\cite{Becker2016b}, have already been the target of research to enhance compiler error messages, while we are not aware of any similar effort for Python.

Our plugin \textsc{Pycee} parses a Python compiler error message automatically as soon as a programmer encounters an error. \textsc{Pycee} then constructs a Stack Overflow query and uses query expansion and reformulation depending on the error type, e.g., by adding related verbs and syntax from other programming languages to the query. After selecting an answer from Stack Overflow, \textsc{Pycee} produces a customised summary of the text and code fragments in the answer and displays the resulting enhanced compiler error message in the Sublime Text IDE.

We evaluated \textsc{Pycee} through a think-aloud user study during which 16 participants---fourteen professionals and two students---completed programming tasks while using two variants of \textsc{Pycee}: the original \textsc{Pycee} which retrieves its data from Stack Overflow and \textsc{PyceeDoc} which retrieves its data from the official Python documentation. In total, our participants encountered 115 compiler errors during the study. The majority of our participants agreed that \textsc{Pycee} was helpful while completing the study tasks, citing concrete suggestions for fixes and example code as major benefits. \textsc{Pycee} generally outperformed \textsc{PyceeDoc} in terms of perceived helpfulness. Our findings confirm that external knowledge sources, such as Stack Overflow, are not only helpful as reference documents, but they can also be harvested automatically to enhance compiler error messages inside an IDE. \textsc{Pycee} is available as an open-source project on GitHub.\footnote{\url{https://github.com/EmillieT/Pycee}\label{pycee-github}}

\section{Motivating Examples}

In this section, we present two examples of \textsc{Pycee} in action, taken from our user study (cf.~Section~\ref{sec:method}).

\paragraph{Invalid syntax}

Figure~\ref{fig:pycee} shows a screenshot of one of our study participants using \textsc{Pycee}. As shown in the Figure, the participant was not familiar with Python syntax and wrote \pythoninline{else if} instead of \pythoninline{elif} in a conditional statement aimed at implementing Rock-paper-scissors.\footnote{Note that the \pythoninline{else if} instance in line 16 had already been corrected by the time the screenshot was taken---line 19 shows an uncorrected one.} The Python compiler responded with \compilerinline{SyntaxError: invalid syntax} (first four lines on white background in Figure~\ref{fig:pycee}), but did not provide any concrete help on how to solve the error. The remaining lines on white background in Figure~\ref{fig:pycee} show \textsc{Pycee}'s output in this scenario, clearly stating how to fix the error (\compilerinline{In python "else if" is spelled "elif"}) and providing a code fragment as example. The output produced by \textsc{Pycee} is a summary of Stack Overflow answer 2395167,\footnote{\url{https://stackoverflow.com/a/2395167}} containing a subset of the answer's sentences as well as a part of the answer's code fragment.

\paragraph{`list' object is not callable}

The following two code fragments show another example of a compiler error encountered by one of our participants who was working on a function to deduplicate lists. The compiler detected a \compilerinline{TypeError}, explaining that \compilerinline{'list' object is not callable}. 

\begin{python}
list = [3, 3, 5, 7, 7, 9, 11, 11]
new_list = list(dict.fromkeys(list))

print(new_list)
\end{python}

\begin{compiler}
Running python -u "/Users/.../2-14.py"
From compiler:
Traceback (most recent call last):
  File "/Users/.../2-14.py", line 18, in <module>
    new_list = list(dict.fromkeys(list))
TypeError: 'list' object is not callable
\end{compiler}

\textsc{Pycee} used text and code from Stack Overflow answer 12836173\footnote{\url{https://stackoverflow.com/a/12836173}} to enhance this message, recommending to not \compilerinline{use tuple, list or other special names as a variable name}, along with an example where \pythoninline{list} had been replaced with \pythoninline{l}:

\begin{compiler}
It should work fine.
Don't use tuple, list or other special names as
  a variable name.
It's probably what's causing your problem.
>>> l = [4,5,6]
>>> tuple(l)
(4, 5, 6)
\end{compiler}

In this case, \textsc{Pycee} did not summarise the answer since it contains fewer than five sentences. In the next section, we describe how \textsc{Pycee} works.

\section{\textsc{Pycee}}
\label{sec:pycee}

\textsc{Pycee} works in two phases: In the first phase, the Python compiler error message is parsed and used to construct a query for Stack Overflow. In the second phase, an answer from Stack Overflow is selected, customised, and summarised. We describe the details of this process in this section.

\subsection{Compiler Error Message Parsing and Query Construction}

As soon as the user encounters a Python compiler error while working in the Sublime Text IDE, \textsc{Pycee} parses the corresponding compiler error message and determines the name of the affected Python file as well as the libraries which the user had imported, the error type (e.g., \compilerinline{SyntaxError}), and whether the compiler returned a specific error message (e.g., \compilerinline{EOL while scanning string literal}). \textsc{Pycee} then constructs a Stack Overflow query, with the exact content of the query depending on the type of compiler error. The following settings were used in the final evaluation described in Section~\ref{sec:method}, and they are the result of extensive experimentation and a preliminary evaluation with student participants. Note that the current implementation of \textsc{Pycee} queries Stack Overflow every time an error is raised. Caching responses to common errors is part of our future work.

\begin{description}[leftmargin=0cm]

\item[AttributeError, NameError.] In the case of an \compilerinline{AttributeError} or a \compilerinline{NameError}, \textsc{Pycee} uses ideas from query reformulation~\cite{Haiduc2013} and query expansion~\cite{Lu2015} to construct a suitable Stack Overflow query. \textsc{Pycee} first extracts the word(s) identified by the compiler as problematic, i.e., the ones surrounded by single quotation marks (e.g., \compilerinline{module} and \compilerinline{Number} in the error message \compilerinline{`module' object has no attribute `Number'}). Following Stefik and Siebert's conjecture that syntactical variations of programming language constructs might affect accuracy among programmers~\cite{Stefik2013}, \textsc{Pycee} then looks up each word in an online resource cataloguing syntax across programming languages.\footnote{\url{http://rigaux.org/language-study/syntax-across-languages.html}} \textsc{Pycee} attempts to locate each word in one of the tables in this resource, and if successful, replaces the word with the most frequently occurring word in the same table while removing any non-letters. Adopting the insights of related work with regard to the importance of tasks (e.g.,~\cite{Kersten2006})---programmers are usually not interested in a concept by itself, but work with the concept as part of an action or task---\textsc{Pycee} then attempts to find actions associated with each word. We use task phrases extracted by TaskNav~\cite{Treude2015a, Treude2015b} from the titles of the one million most recent Stack Overflow threads tagged with \compilerinline{python} as input data, and determine the verb most commonly associated with each word. In addition, the following Python data types are replaced by their English equivalent (e.g., \pythoninline{str} $\rightarrow{}$String) during query construction: \pythoninline{int}, \pythoninline{bool}, \pythoninline{str}, and \pythoninline{dict}.\footnote{Note that the names of other datatypes which can be retrieved using the \pythoninline{dir(__builtins__)} command, such as \pythoninline{tuple}, are already English words.} Words are replaced with their most common domain-specific synonym, using the SEWordSim database~\cite{Tian2014}. The final search query is then expanded to contain the error type, the words, and the associated verbs.

\item[SyntaxError.] In the case of a \compilerinline{SyntaxError}, \textsc{Pycee} first determines whether the error might stem from a common Python programming mistake~\cite{Guo2013}, i.e., mismatched quotes, mismatched brackets, or incorrect syntax for for-loops, while-loops, and conditionals. In these cases, \textsc{Pycee} adds the terms \compilerinline{quotation marks}, \compilerinline{bracket meanings}, \compilerinline{for loop}, \compilerinline{while loop}, and \compilerinline{else if syntax} to the query, respectively. Using a catalogue of Python keywords and builtins, \textsc{Pycee} attempts to fix typos if the word similarity is at least 0.6 using Python's \compilerinline{get_close_matches} function.\footnote{\url{https://docs.python.org/2/library/difflib.html\#difflib.get_close_matches}} If the error was not caused by a common mistake, the search query is \compilerinline{SyntaxError: invalid syntax}.

\item[TypeError.] In the case of a \compilerinline{TypeError}, \textsc{Pycee} uses the phrase \compilerinline{must have first callable argument} as Stack Overflow query if the compiler error message contains the words \compilerinline{the first argument must be callable}, and it removes the error type from the query if the compiler error message contains the phrase \compilerinline{not all arguments converted during string formatting}. These two exceptions were added as a result of our preliminary user study. In all other cases, the original error message is used for the query, including error type and description.

\item[IndentationError, TabError.] In the case of an \compilerinline{IndentationError} or a \compilerinline{TabError}, \textsc{Pycee} performs a Stack Overflow query with the error description only, i.e., not including the error type.

\item[KeyError.] In the case of a \compilerinline{KeyError}, \textsc{Pycee} queries Stack Overflow using only the error type, but not its description.

\item[All other cases.] For all other cases, \textsc{Pycee} queries Stack Overflow using the error type and its description. 

\end{description}

We do not include code fragments as part of the Stack Overflow queries since past work has shown that the Stack Overflow search does not handle code fragments well~\cite{Monperrus2014}. We encourage readers to inspect \textsc{Pycee}'s source code\textsuperscript{\ref{pycee-github}} for additional details.

\subsection{Answer Selection, Customisation, and Summarisation}

The Stack Overflow query is configured to only return threads tagged with \compilerinline{python} which contain at least one answer, sorted by relevance according to the Stack Overflow search algorithm.\footnote{\url{https://meta.stackoverflow.com/questions/355532/how-does-sort-by-relevance-work}} \textsc{Pycee} considers the first page of search results, i.e., up to ten Stack Overflow threads, and selects the first accepted answer in these ten threads for further processing. If no answer has been accepted, the answer with the highest score is selected, provided its score is greater than zero. If no such answer is available, \textsc{Pycee} does not produce an enhanced compiler error message.

The selected answer is customised and summarised as follows:

\begin{description}[leftmargin=0cm]

\item[Customisation.] \textsc{Pycee} locates error messages in the answer's code fragments, identifies the line which caused the error, and replaces any error message in the answer's code with the compiler error message encountered by the user to better fit the code to the user's situation. If the error is a \compilerinline{SyntaxError}, \textsc{Pycee} uses the arrow (\compilerinline{^}) position in the compiler error message to verify that the compiler has identified the correct error line. If that is not the case and if the Stack Overflow answer mentions an offending line (the erroneous line), \textsc{Pycee} replaces this line with the selected line from the answer.

\item[Summarisation.] \textsc{Pycee} summarises the selected answer using Luhn's summarisation algorithm~\cite{Luhn1958} to at most four sentences, following the advice of Nienaltowski et al.~\cite{Nienaltowski2008} who reported that longer error messages do not benefit programmers. We experimented with different summarisation algorithms, aiming to maximise the quality criteria of understandability, completeness, and efficiency~\cite{Moreno2013}, and found Luhn's algorithms to work best on our data. Special characters are converted into a more user-friendly format (e.g., \compilerinline{&gt} $\rightarrow{}$\compilerinline{>}), and unnecessary formatting and padding are removed from the summary. Note that while the goal of the summarisation is to extract parts of posts which are most relevant to an error, there is of course no guarantee that the summary will only contain relevant content.

\end{description}

\textsc{Pycee} then returns the result and displays it below the original compiler error message in the Sublime Text IDE, cf.~Figure~\ref{fig:pycee}.

\section{Evaluation Methodology}
\label{sec:method}

In this section, we outline our evaluation methodology for \textsc{Pycee} in terms of research questions and data collection and analysis.

\subsection{Research Questions}

While some researchers have identified common problems with compiler error messages (e.g.,~\cite{Traver2010, Becker2016a, Marceau2011, Becker2016b}), not much of this work has focused on the Python programming language, with Guo's work~\cite{Guo2013} as a notable exception. However, the focus of Guo's work is not on programmer perceptions of compiler error messages. Therefore, and to establish a baseline for \textsc{Pycee}, with our first research questions, we ask:

\begin{description}
\item[RQ1] How do programmers perceive Python compiler error messages?
\end{description}

After establishing this baseline, our next research question analogously investigates the compiler error messages produced by \textsc{Pycee}:

\begin{description}
\item[RQ2] How do programmers perceive working with \textsc{Pycee}?
\end{description}

One of the key features of \textsc{Pycee} is its reliance on Stack Overflow as a data source for enhancing compiler error messages. To investigate the impact of this feature, we compare \textsc{Pycee} to a baseline with similar functionality but with the official Python documentation as data source. We conduct this comparison in terms of perceived helpfulness, perceived potential time savings, and programmer preferences, as captured by our last research question:

\begin{description}
\item[RQ3] How do programmers perceive the two \textsc{Pycee} variants?
\begin{description}
\item[RQ3.1] How do programmers perceive the helpfulness of the \textsc{Pycee} variants?
\item[RQ3.2] How do programmers perceive the potential time savings of the \textsc{Pycee} variants?
\item[RQ3.3] Which \textsc{Pycee} variant do programmers prefer and why?
\end{description}
\end{description}

\subsection{Data Collection}

To answer our research questions, we conducted a think-aloud user study during which 16 participants---fourteen professional programmers and two students---used two \textsc{Pycee} variants while completing programming tasks in Python. In the following, we characterise the study protocol, tasks, and participants, and we introduce the baseline tool \textsc{PyceeDoc}.

\subsubsection{Study Protocol}

\begin{table*}
\centering
\caption{Questions asked during the user study}
\begin{tabular}{lll}
\toprule
& \textbf{question} & \textbf{answer} \\
\midrule
\parbox[t]{2.5mm}{\multirow{7}{*}{\rotatebox[origin=c]{90}{Before}}} & How many years programming experience do you have? & int \\
 & Do you use programming for your job? If so, what do you do? & description \\
 & How much experience do you have with Python? & int, description \\
 & Do you believe Python compiler error messages provide enough information? & yes/no, description \\
 & What resources do you usually reference when debugging code? & description \\
 & Do you believe the information in these resources usually assists you in solving your problem? & yes/no, description \\
 & What do you usually use to edit code? & list \\
\midrule
\parbox[t]{2.5mm}{\multirow{8}{*}{\rotatebox[origin=c]{90}{During}}} & What are you doing / thinking now? (prompt if required) & extended answer \\
 & Overall, do you believe the plugin was helpful? & agreement scale \\
 & Why? & extended answer \\
 & Overall, do you believe the plugin saved time? & agreement scale \\
 & Why? & extended answer \\
 & Overall, were you satisfied with the information provided by the plugin? & agreement scale \\
 & Why? & extended answer \\
 & What did you like and/or dislike about the plugin? & extended answer \\
\midrule
\parbox[t]{2.5mm}{\multirow{5}{*}{\rotatebox[origin=c]{90}{After}}} & Which plugin did you prefer? & \textsc{Pycee}/\textsc{PyceeDoc} \\
 & Why did you prefer this plugin? & extended answer \\
 & What could make this plugin better? & extended answer \\
 & What would the ideal debugging plugin/tool do? & extended answer \\
 & Any other comments? & extended answer \\
\bottomrule
\vspace{-0.5cm}
\end{tabular}
\label{tab:questions}
\end{table*}

Table~\ref{tab:questions} shows the questions that we asked each participant before, during, and after the study. To ensure that all participants had the same experience and used the same versions of Sublime Text and Python as well as the same set of pre-installed libraries, all participants used the Sublime Text IDE plugins \textsc{Pycee} and \textsc{PyceeDoc} by remotely connecting to the first author's machine using the TeamViewer remote access software.\footnote{\url{https://www.teamviewer.com/en/}}

All studies started with questions aimed at collecting demographic information regarding programming experience, programming job, preferred code editor, and Python experience, followed by questions aimed at answering our first research question about programmers' perceptions of Python compiler error messages and their usual debugging processes.

Participants were then given at least one Python task (see below for task selection) to solve with each variant of \textsc{Pycee} enabled, with the order in which the variants were chosen alternating between participants. Participants were generally limited to 20 minutes per task, however, if they were willing to spend more than one hour for the overall study session, they were allowed to spend more time on each individual task. If a participant had not written any code five minutes into a task or verbally expressed that they had no idea what the task was asking them to do or how to solve it, they were offered another task. If a participant did not encounter any compiler error while completing a task, they were given another task until they encountered a compiler error to ensure that both \textsc{Pycee} variants were used by all participants. Depending on how long participants took per task, they were given the choice of another task or a chance to improve their existing solution until the end of the session.

To best replicate a real programming setting, participants were explicitly given permission to execute code and to refer to the Internet while completing the study tasks. We clarified that the purpose of the study was not to assess their programming ability and we encouraged them to think aloud during the study. To not bias participants with leading questions or hints, we only prompted them by asking the pre-defined questions ``What are you doing / thinking now?'' during their work if needed.

After participants had used a \textsc{Pycee} variant, we inquired about its perceived helpfulness, time savings, satisfaction, and preferences along with the corresponding reasons. To ensure that responses to these questions were unbiased, participants were told no information about the \textsc{Pycee} variants other than they provide ``enhanced error messages''. At the end of each study session, we asked which \textsc{Pycee} variant they preferred and how the tool could be further improved.

All study sessions were video and audio recorded, and extensive notes were taken during each session. \textsc{Pycee} recorded all compiler errors encountered during a session and stored information regarding the error type and description. Note that we cannot make raw data available to protect our participants' privacy.

\subsubsection{Study Tasks}

\begin{table}
\centering
\caption{Assignment of tasks to participants, stars indicate difficulty. $P_{6}$ and $P_{7}$ skipped one task each. $P_{13}$ skipped two tasks and preferred to execute code for their own task for \textsc{PyceeDoc}.}
\label{tab:tasks}
\begin{tabular}{l@{\hskip .3cm}l@{\hskip .3cm}l}
\toprule
 & \textsc{PyceeDoc}                      & \textsc{Pycee} \\
\midrule
$P_{1}$          & Password Generator****        & Decode A Web Page**** \\
                 &                               & Decode A Web Page Two**** \\
$P_{2}$          & Password Generator****        & Decode A Web Page**** \\
                 &                               & Decode A Web Page Two**** \\
$P_{3}$          & Password Generator****        & Decode A Web Page**** \\
$P_{4}$          & Decode A Web Page****         & Password Generator**** \\
$P_{5}$          & Tic Tac Toe Game***           & Decode A Web Page**** \\
                 &                               & Rock Paper Scissors*** \\
$P_{6}$          & Check Primality Functions***  & Reverse Word Order*** \\
$P_{7}$          & Check Primality Functions***  & Tic Tac Toe Game*** \\
$P_{8}$          & Guessing Game One***          & Cows And Bulls*** \\
$P_{9}$          & Guessing Game One***          & Password Generator**** \\
$P_{10}$         & Rock Paper Scissors***        & Guessing Game One*** \\
$P_{11}$         & Tic Tac Toe Game***           & Guessing Game One*** \\
$P_{12}$         & Guessing Game Two***          & Cows And Bulls*** \\
$P_{13}$         & \textit{Own}                  & Guessing Game One*** \\
$P_{14}$         & List Overlap Comprehensions** & Fibonacci** \\
                 &                               & List Remove Duplicates** \\
$P_{15}$         & Tic Tac Toe Game***           & Cows And Bulls*** \\
$P_{16}$         & List Comprehensions**         & String Lists** \\
\bottomrule
\vspace{-0.5cm}
\end{tabular}
\end{table}

To use objective criteria for the creation of study tasks to the extent possible and not bias the tasks toward \textsc{Pycee}, we chose study tasks from the Practice Python website, a resource aimed at providing small, short, and relevant introductory Python programming exercises for beginners.\footnote{\url{https://www.practicepython.org/}} The site contains 25 Python tasks after merging sub-tasks which are part of a larger task and excluding tasks not suitable for the study, categorised by their difficulty into four categories (from one chilli up to four chillies, with increasing difficulty). We eventually excluded two tasks which required stable and fast Internet access (decoding a web page, parts one and two) after we noticed that the Internet connection speed of participants varied. We also excluded one task that required graphing of data, which did not work in our setup with Sublime Text. To ensure that participants did not find the tasks too trivial and to give us a realistic chance at encountering compiler errors, tasks were allocated to participants according to their difficulty rating: For participants who indicated at most half a year of Python experience, we allocated two-chilli tasks, and for participants who indicated at least half a year of Python experience, we allocated three- and four-chilli tasks. Within these constraints, tasks were assigned randomly to participants. Table~\ref{tab:tasks} shows the assignment.

\subsubsection{Study Participants}

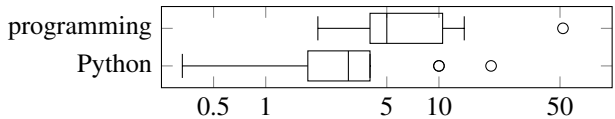
\begin{figure}
\centering
\begin{tikzpicture}
\begin{axis}[y=0.5cm, x=1cm, try min ticks=2, ytick=5, xmin=0.25, xmax=100, xmode=log, xtick={0.5,1,5,10,50}, xticklabels={0.5,1,5,10,50}, ytick={1,2}, yticklabels={Python, programming}]
\addplot[black, mark=o, boxplot, color=black]
table[row sep=\\,y index=0] {
data\\
0.33\\ 0.5\\ 1\\ 1\\ 2\\ 2\\ 2.5\\ 3\\ 3\\ 3\\ 4\\ 4\\ 4\\ 10\\ 10\\ 20\\
};
\addplot[black, mark=o, boxplot, color=black]
table[row sep=\\,y index=0] {
data\\
2\\ 3\\ 3\\ 4\\ 4\\ 4\\ 4\\ 5\\ 5\\ 6\\ 8\\ 10\\ 12\\ 14\\ 14\\ 52\\ 
};
\end{axis}
\end{tikzpicture}
\caption{Participant experience in years (log scale)}
\label{fig:experience}
\end{figure}

\begin{table}
\centering
\caption{Resources used when debugging code}
\begin{tabular}{lr}
\toprule
\textbf{resource} & \textbf{participants} \\
\midrule
Stack Overflow & 13 \\ 
Google & 10 \\
Official Documentation & 6 \\
Compiler & 1 \\
\bottomrule
\vspace{-0.5cm}
\end{tabular}
\label{tab:resources}
\end{table}

\begin{table}
\centering
\vspace{-0.5cm}
\caption{Code editors usually used by participants}
\begin{tabular}{lr}
\toprule
\textbf{resource} & \textbf{participants} \\
\midrule
Sublime Text & 7 \\ 
PyCharm & 6 \\
Visual Studio & 4 \\
Vim & 2 \\
\bottomrule
\vspace{-0.8cm}
\end{tabular}
\label{tab:editors}
\end{table}

We advertised our study on social media and through professional contacts and we recruited participants through the freelancing website Upwork,\footnote{\url{https://www.upwork.com/}} resulting in a total of 16 participants. The majority of participants ($\nicefrac{14}{16}=88\%$) were professional programmers who use programming as part of their job, the remaining two were students. Job titles ranged from software developer and project manager to CTO and big data engineer. Figure~\ref{fig:experience} shows the experience of participants in years. Participants had a median of five years of programming experience, with a minimum of two years and a maximum of 52 years, and they had a median of three years of Python experience, with a minimum of four months and a maximum of 20 years.

As shown in Table~\ref{tab:resources}, the majority of participants indicated to resort to Stack Overflow ($\nicefrac{13}{16}=81\%$) and Google ($\nicefrac{10}{16}=63\%$) when debugging code. Other resources mentioned included colleagues, textbooks, tutorial sites, and discussion forums. Table~\ref{tab:editors} shows the code editors usually used by participants, with Sublime Text ($\nicefrac{7}{16}=44\%$) and PyCharm ($\nicefrac{6}{16}=38\%$) being the most common. Other editors mentioned included Eclipse, IntelliJ, Atom, and Vi.

\subsubsection{\textsc{PyceeDoc}}

To be able to assess the helpfulness of using content from Stack Overflow to enhance Python compiler error messages, we implemented a baseline called \textsc{PyceeDoc} which accesses the official Python documentation for each compiler error instead of Stack Overflow. When encountering a compiler error, \textsc{PyceeDoc} reproduces the corresponding content from the Exceptions page of the Python API\footnote{\url{https://docs.python.org/3/library/exceptions.html}} in the Sublime Text IDE. Note that \textsc{PyceeDoc} does not make use of the description of compiler error messages, but only uses the error type since there is only exactly one explanation available in the official documentation for each error type. To be compatible with the Sublime Text IDE, we removed links from the documentation as well as any version changes. As an example, the documentation for \pythoninline{IndentationError} states: ``Base class for syntax errors related to incorrect indentation. This is a subclass of \pythoninline{SyntaxError}''.\footnote{\url{https://docs.python.org/3/library/exceptions.html\#IndentationError}}

\subsection{Data Analysis}

In this section, we describe how we analysed the collected data to answer each of our research questions.

\subsubsection{RQ1: Programmer perceptions of Python compiler error messages}

To answer our first research question, we analysed participant responses about general perceptions of Python error messages and the resources typically referenced when debugging (cf.~Table~\ref{tab:questions}). We conducted open coding following the definition of Strauss and Corbin~\cite{Strauss1998}, i.e., generating categories and considering their variations response by response~\cite{Stol2016}. The coding was done by the second author using NVivio~\cite{Bazeley2013} and verified by the first author. Where applicable, we quote participants when presenting findings to increase traceability to raw data. We show a subset of codes in the following text in \textit{italics}, and we indicate how many participants mentioned the particular code in superscript. Note that these numbers only indicate how much evidence the data analysis yielded for each code, they do not necessarily indicate the importance of a code since we did not explicitly ask all participants about each code specifically.

\subsubsection{RQ2: Programmer perceptions while using \textsc{Pycee}}

To answer our second research question, we analysed the notes we took and the screen and audio recordings of participants using \textsc{Pycee} during the think-aloud part of the study. Since these notes were taken separately for each of the compiler errors encountered by participants during the study, we analysed the notes compiler error by compiler error during open coding. The coding was conducted again by the second author and verified by the first author.

\subsubsection{RQ3: Programmer perceptions of \textsc{Pycee} variants}

To answer our last research question, we analysed the participants' responses regarding helpfulness, time savings, satisfaction, and preference of each \textsc{Pycee} variant. In addition to reporting the responses on the Likert scales, we qualitatively analysed the reasons that participants gave for their opinions, using the previously described qualitative analysis process.

\section{Findings}
\label{sec:findings}

In this section, we describe our findings for each research question.

\subsection{RQ1: Programmer perceptions of Python compiler error messages}

Most participants indicated that Python compiler error messages had room for improvement, citing issues when encountering complex problems and the need to look at other resources. One of the codes which emerged from our qualitative analysis of participants' responses regarding their perception of Python compiler error messages was that they are \textit{bad for beginners}$^{(4)}$. For example, when asked whether Python compiler error messages provide sufficient information, $P_{4}$ explained that ``Only to people who are familiar with the error'' and $P_{8}$ answered ``Generally no but yes after years of experience''.

The majority of participants ($\nicefrac{12}{16}=75\%$) indicated that information in other resources, such as Stack Overflow, usually assists them in solving their problems. However, looking up information in external resources is not trivial, as explained by $P_{15}$: ``Most of the time but can be consuming for unique cases''. These results support our initial motivation in building \textsc{Pycee} to bridge the gap between Python compiler errors and documentation found in external resources.

\begin{mdframed}[style=mystyle,frametitle=Summary RQ1]
The majority of participants perceived Python compiler error messages to have room for improvement, in particular for beginners. External resources could usually assist participants when encountering a compiler error, but this might be time-consuming.
\end{mdframed}

\subsection{RQ2: Programmer perceptions while using \textsc{Pycee}}

\begin{table*}
\centering
\caption{Compiler errors encountered by study participants}
\label{tab:errors}
\begin{tabular}{ll|rr|rr}
\toprule
 &  & \multicolumn{2}{c|}{\textsc{PyceeDoc}} & \multicolumn{2}{c}{\textsc{Pycee}} \\
\textbf{type} & \textbf{description} & \textbf{occurrences} & \textbf{by type} & \textbf{occurrences} & \textbf{by type}\\
\midrule
AttributeError & `X' object has no attribute `Y' & 4 & 4 & 6 & 6\\
\midrule
ImportError & cannot import name `X' & 1 & 4 & -- & 2\\
 & No module named `X' & 3 &  & 2 & \\
\midrule
IndentationError & expected an indented block & 1 & 3 & 8 & 10\\
 & unindent does not match any outer indentation level & 2 &  & 2 & \\
\midrule
IndexError & list index out of range & -- & 3 & 2 & 2\\
 & `X' index out of range & 3 &  & -- & \\
\midrule
KeyError & `class' & -- & -- & 1 & 1\\
\midrule
NameError & global name `X' is not defined & 1 & 11 & 1 & 13\\
 & name `X' is not defined & 10 &  & 12 & \\
\midrule
SyntaxError & EOL while scanning string literal & 1 & 19 & 1 & 21\\
 & invalid syntax & 18 &  & 20 & \\
\midrule
TypeError & can only concatenate list (not ``str'') to list & 1 & 7 & -- & 5\\
 & cannot concatenate `str' and `int' objects & 1 &  & 1 & \\
 & `int' object is not iterable & 1 &  & 2 & \\
 & `list' object is not callable & -- &  & 1 & \\
 & `NoneType' object is not callable & 2 &  & -- & \\
 & unsupported operand type(s) for +: `X' and `Y' & 1 &  & -- & \\
 & `X' takes exactly `Y' arguments (`Z' given) & 1 &  & 1 & \\
\midrule
ValueError & invalid literal for int() with base 10: `X' & -- & -- & 1 & 1\\
\midrule
ZeroDivisionError & integer division or modulo by zero & 2 & 2 & 1 & 1\\
\midrule
\textbf{sum} &  & 53 & 53 & 62 & 62\\
\bottomrule
\vspace{-0.5cm}
\end{tabular}
\end{table*}

Table~\ref{tab:errors} shows the compiler errors encountered by our study participants while using \textsc{Pycee} and its baseline variant \textsc{PyceeDoc}. In total, participants encountered 115 compiler errors, 62 while working with \textsc{Pycee} and 53 while working with \textsc{PyceeDoc}. The most common error type was \compilerinline{SyntaxError} followed by \compilerinline{NameError}. All participants encountered at least one compiler error per \textsc{Pycee} variant with a median number of two and a half errors while using \textsc{PyceeDoc} and four errors while using \textsc{Pycee}. Error messages which only differed in their variable names have been grouped and errors caused by Sublime Text or \textsc{Pycee}, such as connection errors, have been filtered out.

When using the baseline tool \textsc{PyceeDoc}, several participants criticised that the enhanced compiler error message---directly copied from the Python API documentation---contained \textit{too much information}$^{(6)}$. $P_{9}$ for example encountered a \compilerinline{TypeError} and did not bother to read all information provided by the tool, stating ``It's a bit too long''. Another common complaint was that the output of \textsc{PyceeDoc} was \textit{too generic}$^{(6)}$, e.g., $P_{1}$: ``It did not tell me what I should be doing''. Participants also noted that the documentation was \textit{too formal}$^{(3)}$, e.g., $P_{3}$ noted ``This is not clear at all, I want plain English'' after encountering an \compilerinline{IndentationError}. There was a general feeling among participants that a tool for enhancing compiler error messages should focus on common errors, as stated by $P_{3}$: ``You should use common (basic) errors as test cases when testing this plugin''.

When using \textsc{Pycee}, participants commended the tool for \textit{suggesting a fix or giving an example}$^{(4)}$ in various scenarios. For example, $P_{14}$, after encountering a \compilerinline{SyntaxError}, stated ``It's helpful and suggests how to fix the problem'' and $P_{12}$, also after encountering a \compilerinline{SyntaxError}, added ``I like the language and the code examples''. These comments show that at least in some scenarios, \textsc{Pycee} was able to address the main weaknesses participants perceived when working with the baseline tool: a lack of specificity, formal language, and too much information. Participants further noted the additional context provided by \textsc{Pycee}, as expressed by $P_{4}$: ``I liked the context and that it explains the specific location''. To some extent, the success of \textsc{Pycee} \textit{depended on the type of error}$^{(3)}$, as noted by $P_{5}$: ``It has useful information but not for this case [a \compilerinline{TypeError}]''. In some cases, the information provided by \textsc{Pycee} was \textit{incorrect}$^{(3)}$. For example, $P_{1}$ encountered a \compilerinline{NameError} and stated: ``It provided me with the wrong information, it was a global variable but I was told to look in local variables''.

\begin{mdframed}[style=mystyle,frametitle=Summary RQ2]
During the think-aloud part of the study, participants encountered a total of 115 enhanced compiler error messages. Common perceptions about the compiler error messages generated by the baseline tool \textsc{PyceeDoc} were that these messages contained too much information and were too generic. Compiler error messages generated by \textsc{Pycee} were commended for including suggestions for fixes and examples, but were not perceived to be correct in all cases.
\end{mdframed}

\subsection{RQ3: Programmer perceptions of \textsc{Pycee} variants}

\begin{figure}
\centering
\begin{tikzpicture}
\begin{axis}[y=0.5cm, x=1cm, try min ticks=2, ytick=5, xmin=-0.5, xmax=4.5, xtick={0,4}, xticklabels={strongly disagree, strongly agree}, ytick={1,2}, yticklabels={\textsc{Pycee}, \textsc{PyceeDoc}}]
\addplot[black, mark=o, boxplot prepared={%
median=3,
upper quartile=3.75,
lower whisker=1.875,
lower quartile=3,
upper whisker=4
}, color=black] table [row sep=\\,y index=0] { 0\\ 0\\ };
\addplot[black, mark=o, boxplot prepared={%
median=3,
upper quartile=3,
lower whisker=0,
lower quartile=1.25,
upper whisker=4
}, color=black] coordinates {};
\end{axis}
\end{tikzpicture}
\caption{Perceived helpfulness of \textsc{Pycee} variants}
\vspace{-0.5cm}
\label{fig:helpful}
\end{figure}
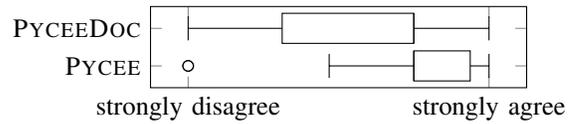

After completing the think-aloud portion of the study, we asked participants about their perceptions of the \textsc{Pycee} variants in terms of helpfulness, time savings, and satisfaction, along with their preference for one variant over the other.

Figure~\ref{fig:helpful} shows the results for helpfulness on a 5-point Likert scale. All but three participants (i.e., $\nicefrac{13}{16}=81\%$) agreed or strongly agreed with \textsc{Pycee}'s helpfulness. The number for the baseline tool \textsc{PyceeDoc} is slightly lower at $\nicefrac{11}{16}=69\%$. For \textsc{Pycee}, participants explained their rating with \textit{successful instances}$^{(4)}$ where the tool had worked well, e.g., $P_{16}$ stated: ``I liked how the \compilerinline{NameError} message showed how to check for the error''. For \textsc{PyceeDoc}, participants pointed at \textit{novices as a target audience}$^{(4)}$, e.g., $P_{2}$ explained: ``It wasn't helpful for me but it would be for novices''.

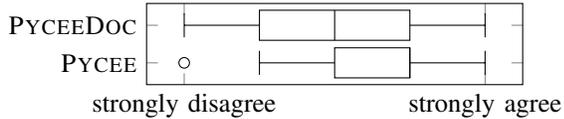
\begin{figure}
\centering
\begin{tikzpicture}
\begin{axis}[y=0.5cm, x=1cm, try min ticks=2, ytick=5, xmin=-0.5, xmax=4.5,xtick={0,4}, xticklabels={strongly disagree, strongly agree}, ytick={1,2}, yticklabels={\textsc{Pycee}, \textsc{PyceeDoc}}]
\addplot[black, mark=o, boxplot, color=black]
table[row sep=\\,y index=0] {
data\\
0\\ 1\\ 1\\	2\\ 2\\ 2\\	3\\	3\\	3\\	3\\	3\\	3\\	3\\	3\\	4\\ 4\\
};
\addplot[black, mark=o, boxplot, color=black]
table[row sep=\\,y index=0] {
data\\
0\\ 0\\ 1\\ 1\\ 1\\ 1\\ 2\\ 2\\ 2\\ 2\\ 3\\ 3\\ 3\\ 3\\ 4\\ 4\\
};
\end{axis}
\end{tikzpicture}
\caption{Perceived time savings of \textsc{Pycee} variants}
\vspace{-0.5cm}
\label{fig:timesavings}
\end{figure}

As shown in Figure~\ref{fig:timesavings}, there were no discernible differences in the distributions of participant perceptions regarding the potential for time savings between the two \textsc{Pycee} variants. $\nicefrac{10}{16}=63\%$ participants for \textsc{Pycee} and $\nicefrac{6}{16}=38\%$ participants for \textsc{PyceeDoc} agreed or strongly agreed with the statement that the tool saved time. Note that these results might be biased against the \textsc{Pycee} variants due to slow Internet connection of some participants as a result of the remote study setup. The reasons which participants mentioned for their perceptions did not differ much between the \textsc{Pycee} variants: Some participants found that they \textit{did not have to look for information elsewhere}$^{(4)}$, such as $P_{16}$ who stated ``Yes, I could skip using the browser''. In other cases where \textsc{Pycee} was not helpful, participants still needed to search elsewhere, e.g., $P_{15}$: ``I still need to search how to fix the problem''.

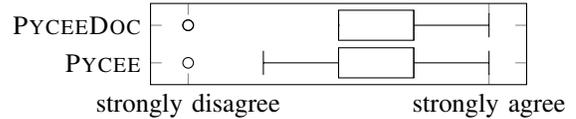
\begin{figure}
\centering
\begin{tikzpicture}
\begin{axis}[y=0.5cm, x=1cm, try min ticks=2, ytick=5, xmin=-0.5, xmax=4.5, xtick={0,4}, xticklabels={strongly disagree, strongly agree}, ytick={1,2}, yticklabels={\textsc{Pycee}, \textsc{PyceeDoc}}]
\addplot[black, mark=o, boxplot, color=black]
table[row sep=\\,y index=0] {
data\\
0\\ 1\\ 1\\ 2\\ 2\\ 2\\ 2\\ 3\\ 3\\ 3\\ 3\\ 3\\ 3\\ 4\\ 4\\ 4\\
};
\addplot[black, mark=o, boxplot, color=black]
table[row sep=\\,y index=0] {
data\\
0\\ 0\\ 2\\ 2\\ 2\\ 2\\ 2\\ 3\\ 3\\ 3\\ 3\\ 3\\ 3\\ 3\\ 4\\ 4\\
};
\end{axis}
\end{tikzpicture}
\caption{User satisfaction for \textsc{Pycee} variants}
\vspace{-0.5cm}
\label{fig:satisfaction}
\end{figure}

In terms of user satisfaction, both \textsc{Pycee} variants scored equally well as shown in Figure~\ref{fig:satisfaction}. In both cases, $\nicefrac{9}{16}=56\%$ agreed or strongly agreed with the statement that they were satisfied with the tool. For \textsc{Pycee}, participants justified their answers by referring to the \textit{style of the enhanced error messages}$^{(4)}$ and the \textit{presence of code examples}$^{(4)}$. For example, $P_{4}$ explained: ``I liked the code examples and full sentences in normal English, written for humans''. On the other hand, one of the disadvantages of \textsc{Pycee} is that it relies on information from Stack Overflow which may or may not be correct. Several participants mentioned this trade-off, e.g., $P_{9}$: ``The additional information is helpful and saves time but the information was not always correct and then you lose time''. For \textsc{PyceeDoc}, participants commended the \textit{additional context}$^{(5)}$ added in the enhanced compiler error message, e.g., $P_{2}$ stated ``Some of it expands the information and provides context ... and focuses on the problem''. However, participants complained about a \textit{lack of direction}$^{(4)}$ in fixing errors, e.g., $P_{13}$: ``It says what has happened but not how to fix it''. 

Finally, we asked participants which \textsc{Pycee} variant they preferred. As shown in Table~\ref{tab:preferred}, there is a preference toward \textsc{Pycee}, with an equally large number of participants not having a preference. Among those who preferred \textsc{Pycee}, i.e., the variant relying on Stack Overflow, participants mentioned reasons such as ``It felt more personal'' ($P_{4}$) and the \textit{presence of examples}$^{(4)}$, e.g., $P_{9}$: ``Both seemed the same, the examples from the plugin were the best part''. Note that all examples came from Stack Overflow and were therefore only available in \textsc{Pycee}.

\begin{table}
\centering
\caption{Preferred \textsc{Pycee} variant}
\begin{tabular}{lr}
\toprule
\textbf{variant} & \textbf{participants} \\
\midrule
\textsc{Pycee} & 7 \\
\textsc{PyceeDoc} & 2 \\ 
no preference & 7 \\
\bottomrule
\vspace{-0.5cm}
\end{tabular}
\label{tab:preferred}
\end{table}

\begin{mdframed}[style=mystyle,frametitle=Summary RQ3]
The majority of participants agreed that \textsc{Pycee} is helpful and that it saves time. When compared to the baseline \textsc{PyceeDoc}, participants generally preferred \textsc{Pycee} in terms of helpfulness, referring to concrete suggestions on how to fix compiler errors and code examples as strong points.
\end{mdframed}

\section{Discussion and Open Research Challenges}
\label{sec:discussion}

Our work has provided evidence that it is indeed possible to use data from online sources such as Stack Overflow to automatically enhance compiler error messages. The majority of participants agreed that \textsc{Pycee} is helpful and that it saves time, primarily thanks to the inclusion of code examples and concrete suggestions on how to fix errors in the automatically generated compiler error messages. A trade-off we encountered as part of this work is the potentially low quality of content on Stack Overflow (see for example Ragkhitwetsagul et al.'s recent work on toxic code snippets on Stack Overflow~\cite{ragkhitwetsagul2019}). While Stack Overflow likely contains information on most errors, not all of it is correct or relevant. In one case, $P_{3}$ had misspelled the Python print command as \pythoninline{pint}. \textsc{Pycee} provided advice on how to check for the existence of a local variable, using the variable name \pythoninline{myVar} in a code example, which led to additional confusion. In another case, $P_{13}$ attempted to concatenate a string and an integer using the \pythoninline{+} symbol. \textsc{Pycee} produced a code snippet with a call to Python's \pythoninline{lambda} function which had no relevance to the solution (a call Python's \pythoninline{str} function). We will continue to explore these challenges in future work, e.g., by trying to identify minimal working examples on Stack Overflow.

With one of the last questions in our study, we asked participants to describe their hypothetical ideal debugging tool. While $P_{11}$'s response sums up the general sentiment well: ``It would solve the error for me so I don't have to do any work'', other participants had more concrete suggestions, such as, a good debugging tool should be helpful in fixing common mistakes ($P_{13}$: ``I would like to see common examples of how the error is thrown and suggested usage of the solution or the function that caused the error''). Current compiler error messages---much like API documentation~\cite{Maalej2013}---focus on covering all possible cases instead of covering the common ones well. Past work has shown that Stack Overflow works the other way around: The crowd is capable of generating a rich source of content with code examples and discussion that is actively viewed and used by many more developers, but does not usually achieve perfect coverage~\cite{Parnin2012}.

Context awareness~\cite{Antunes2014} was mentioned as an important feature of the ideal debugging tool by many participants. Context can refer to relevant links to external resources ($P_{11}$: ``Links to relevant Stack Overflow posts and additional information'') as well as to the user's code base ($P_{12}$: ``Integration of the user's code into the plugin would be great''). We tried to improve \textsc{Pycee}'s context awareness by replacing error messages in the Stack Overflow answer's code with the compiler error message encountered by the user to better fit code examples to the user's situation (cf.~Section~\ref{sec:pycee}), but more work is needed to achieve better context awareness in debugging tools.

Many of our participants discussed the visual appearance of compiler error messages, suggesting that error messages should favour visual content over textual content ($P_{3}$: ``It should have little text and be visual''), exist in a separate layer on top of the source code ($P_{5}$: ``A pop up window with information would be good or when you mouseover the code a message box appears''), be accessible similar to other code elements ($P_{8}$: ``I want it to be more invisible and to be usable with keyboard shortcuts''), and be interactive ($P_{3}$: ``It should be teachable, the user should be able to interact with it''). Not much related work has focused on the human-computer interaction aspect of how to present compiler errors to users (with Barik et al.~\cite{Barik2014} and Prather et al.~\cite{Prather2017} as notable exceptions). Our participant responses suggest that more work is required in this area.

\section{Threats to Validity}
\label{sec:threats}

Similar to other empirical studies, there are threats which may affect the validity of our results.

Threats to the \textit{construct validity} correspond to the appropriateness of the evaluation metrics. We evaluated the \textsc{Pycee} variants in terms of their perceived helpfulness, their perceived potential for time saving, and the user satisfaction. Similar metrics have been used in many other studies before (e.g.,~\cite{Seffah2006}) and these metrics reflect our goals behind developing \textsc{Pycee}. The data on participant experience in Python which we used to allocate programming tasks to participants was derived from participant responses, and we cannot guarantee that these responses accurately reflect each participant's experience.

Threats to the \textit{internal validity} compromise our confidence in establishing a relationship between the independent and dependent variables. Participants were not informed what sources each of the \textsc{Pycee} variants used to ensure that responses would not be biased toward or against one variant. While participants were given the opportunity to use both \textsc{Pycee} variants for at least 20 minutes each, some only encountered as little as one compiler error per variant. These participants would have only experienced a small subset of \textsc{Pycee}'s functionality, missing out on \textsc{Pycee}'s handling of specific error types. In addition, participants did not encounter the same compiler errors while working with the \textsc{Pycee} variants since they solved different tasks with each variant. This would likely have been reflected in their answers. During the study, we noticed that some participants had a tendency to discuss what programmers of other skill levels might think of \textsc{Pycee} instead of focusing on their own programming task. This might have affected their answers, but did likely not affect one variant of \textsc{Pycee} more than the other. One or two participants correctly guessed the sources used by the \textsc{Pycee} variants. We did not confirm their guesses until after the study, but this might still have influenced their answers. Some participants encountered errors which they already knew how to solve, which may have affected the way they perceived the \textsc{Pycee} variants. Since the study was conducted remotely, we cannot guarantee that participants did not search for solutions outside of the study setup.

Threats to \textit{external validity} correspond to the ability to generalise our results. We cannot claim generalisability beyond the Python programming language or the particular implementations of \textsc{Pycee} and \textsc{PyceeDoc} used in our study. Recruiting more or different programmers to participate in the study and asking them to work on different tasks may have led to different results. Note that we decided to give participants freedom in terms of how they tackled their programming tasks to create as realistic a scenario as possible---an alternative design in which participants would have been asked to fix a set of given compiler errors would have been more contrived. Some participants were hesitant to write Python code as they felt their programming abilities were being judged. A small number of participants introduced errors to their code on purpose out of curiosity to trigger \textsc{Pycee} and see its result. These issues might reduce the extent to which our study sessions reflect an actual programming setting.

\section{Related Work}
\label{sec:related}

Work related to \textsc{Pycee} can be grouped into three categories: compiler error message enhancement, use of Stack Overflow content in IDEs, and summarisation in software engineering.

\paragraph{Compiler Error Message Enhancement}

Becker~\cite{Becker2016a} suggested that providing enhanced error messages to novices can reduce the future number of error messages received. While the majority of research uses Java compiler errors, Becker et al.~\cite{Becker2016b} discussed the rising popularity of Python as an introductory programming language, suggesting the need for more research. Nienaltowski et al.~\cite{Nienaltowski2008} found that longer error messages do not necessarily benefit students, and that the additional information provided (e.g., error code) may be a cause for additional confusion. Hristova et al.~\cite{Hristova2003} developed an approach to provide enhanced error messages for Java, focused on enhancing the function of a compiler so that the enhanced message heavily references the users' code.

\paragraph{Use of Stack Overflow Content in the IDE}

Seahawk by Ponzanelli et al.~\cite{Ponzanelli2013} is an Eclipse plugin which automatically formulates queries from the current source code context and presents a ranked and interactive list of Stack Overflow results to the user. A related tool called Prompter was later proposed by the same research group~\cite{Ponzanelli2014}. Cordeiro et al.~\cite{Cordeiro2012} developed a tool which integrates the recommendations of question-and-answer web resources related to stack traces into the Eclipse IDE. AutoComment by Wong et al.~\cite{Wong2013} extracts code-description mappings from Stack Overflow and leverages this information to automatically generate descriptive comments for similar code. NLP2Code by Campbell and Treude~\cite{Treude2017} and Bing Developer Assistant by Zhang et al.~\cite{Zhang2016} provide code snippets in the IDE via natural language queries, while Treude and Robillard~\cite{Treude2017b} found that less than half of Stack Overflow code snippets are considered to be self-explanatory.

\paragraph{Summarisation in Software Engineering}

Haiduc et al.~\cite{Haiduc2010} found that a combination of text summarisation techniques is most appropriate for source code summarisation. Moreno et al.~\cite{Moreno2013} developed an approach to summarise Java classes, McBurney et al.~\cite{McBurney2016} analysed method calls and leveraged the \mbox{PageRank} algorithm to generate a description of the behaviour of a Java method, and Alqaimi et al.~\cite{Alqaimi2019} summarised Java lambda expressions. Ying and Robillard~\cite{Ying2013} developed an approach for the automated summarisation of code fragments, and Buse et al.~\cite{Buse2008} developed an approach to automatically summarise the conditions of Java exceptions. Rastkar et al.~\cite{Rastkar2011} introduced an automated approach which produces a natural language summary describing cross-cutting concerns and how they are implemented. The same research group~\cite{Rastkar2010, Rastkar2014} investigated whether it is possible to summarise bug reports automatically and effectively so that developers can consult summaries instead of entire bug reports.

\section{Conclusions and Future Work}
\label{sec:conclusion}

Motivated by the tendency of compilers to produce cryptic and uninformative error messages and the plethora of Stack Overflow threads discussing compiler errors, we have presented \textsc{Pycee}, a plugin integrated with the Sublime Text IDE to provide enhanced compiler error messages for the Python programming language. \textsc{Pycee} automatically queries Stack Overflow to provide customised and summarised information about compiler errors within the IDE. Our evaluation through a think-aloud study, during which 16 programmers completed programming tasks while using two \textsc{Pycee} variants and encountering a total of 115 compiler errors for which \textsc{Pycee} produced an enhanced compiler error message, showed that the majority of participants agreed that \textsc{Pycee} was helpful while completing the study tasks. Participants primarily cited the concrete suggestions for fixes and code examples included in the enhanced compiler error messages as major benefits of \textsc{Pycee}. Our results confirm that data from online sources such as Stack Overflow can be successfully used to automatically enhance compiler error messages. 

In addition to improving \textsc{Pycee}, in particular in terms of its context awareness and its handling of common mistakes, our future work lies in investigating suitable user interfaces for communicating compiler errors to programmers in order to transform error messages from Wexelblat's foreboding ``cryptic diagnostics''~\cite{Wexelblat1976} into tools which can reliably help programmers solve compiler errors.

\section*{Acknowledgements}

The authors would like to thank Greg Wilson for suggesting to build \textsc{Pycee} and all study participants for their participation. This work was inspired by the International Workshop series on Dynamic Software Documentation, held at McGill's Bellairs Research Institute in February 2017 and February 2018. This work has been supported by the Australian Research Council's Discovery Early Career Researcher Award (DECRA) funding scheme (DE180100153).


\end{sloppy} 
\end{document}